\documentclass[12pt,amsfonts]{article}
\begin{document}
\def\p {{\partial}}
\def\n {{\nu}}
\def\m {{\mu}}
\def\a {{\alpha}}
\def\bt {{\beta}}
\def\f {{\phi}}
\def\th {{\theta}}
\def\g {{\gamma}}
\def\eps {{\epsilon}}
\def\e {{\psi}}
\def\la {{\lambda}}
\def\na {{\nabla}}
\def\k {\chi}
\def\bn {\begin{eqnarray}}
\def\en {\end{eqnarray}}
\title{Quantization of singular systems with second order Lagrangians \footnote{E-mail:
$sami_{-}muslih$@hotmail.com}}
\maketitle
\begin{center}
\author{Sami I. Muslih\\Dept. of Physics\\ Al-Azhar University\\
Gaza, Palestine}
\end{center}
\hskip 5 cm
\begin{abstract}
The path integral formulation of singular systems with second
order Lagrangians is studied by using the canonical path integral
formulation method. The path integral quantization of Podolsky
electrodynamics is studied.
\end{abstract}
\newpage
\section{Introduction}
The study of constrained systems with higher order Lagrangians has
been applied in many relevant physical problems. Poddolsky's
electrodynamics [1] and a relativistic particle with curvature and
torsion in three dimensional space-time [2] are some examples.

The treatment for theories with higher order Lagrangians has been
first developed by Ostrogradski [3 ]and leads to obtain the Euler
and the Hamiltons equations of motion.

The Lagrangian formulation of these theories require the
configuration space formed by $n$ generalized coordinates $q_i,
\dot{q_i}$ and $\ddot{q_i}$. The Euler Lagrangian equations of
motion, which are obtained from
\begin{equation}
S= \int L(q_i, \dot{q_i}, \ddot{q_i})dt,
\end{equation}
using the Hamiltons principle, are given by:
\begin{equation}
\frac{\p L}{\p q_i} -\frac{d}{dt}(\frac{\p L}{\p \dot{q_i}})+
\frac{d^2}{dt^2}(\frac{\p L}{\p \ddot{q_i}})= 0.
\end{equation}

The passage from the Lagrangian approach to the Hamiltonian
approach is achieved by introducing the generalized momenta
$(p_i, \pi_i)$ conjugated to the generalized coordinates $(q_i,
\dot{q_i})$ respectively as

\bn &&p_i=\frac{\p L}{\p \dot{q_i}} -\frac{d}{dt}(\frac{\p L}{\p
\ddot{q_i}}),\\
&&\pi_i =\frac{\p L}{\p \ddot{q_i}}. \en

The phase-space will then be spanned by the canonical variables
$(q_i, p_i)$ and $(\bar{q_i}, \pi_i)$, where
$(\bar{q_i}=\dot{q_i})$.

However, a valid phase space is formed if the rank of the Hessian
matrix
\begin{equation}
\frac{\p^2 L}{\p \ddot{q_i}\p \ddot{q_j}},\;\;i,j= 1,...,n,
\end{equation}
is $ n$. Systems which have this this property are called regular
and their treatments are found in a standard mechanics books.
Systems which have the rank less than $n$ are called singular
systems.

Now we will give two formulations to investigate singular systems
which are: Dirac's method and the canonical path integral
formulation [6-11]
\section{ Dirac method}

The well-known method to investigate the Hamiltonian formulation
of constrained systems was initiated by Dirac [4,5]. In his
formulation one defines the total Hamiltonian as
\begin{equation}
H_T = H_c +\nu_{\a}H'_{\a},\;\;\; \a=1,...,m < 2(n-1),
\end{equation}
where $H_c$ being the canonical Hamiltonian and determined as
\begin{equation}
H_c = p_i \bar{q_i} +\pi_i \ddot{q_i} - L,
\end{equation}
\and $\nu_{\a}$ are unknown coefficients.

Due to the singular nature of the Hessian, we have $\a$
functionally independent relations of the form
\begin{equation}
H'_{\a}(q_i, p_i, \bar{q_i}, \pi_i)\approx 0.
\end{equation}

Consistency conditions
\begin{equation}
\dot{H'_{\a}}= \{H'_{\a} , H_c\} + \nu_{\m} \{H'_{\a},
H'_{\bt}\}\approx 0,
\end{equation}
leads to the secondary constraints . Repeating this procedure as
many times as needed, one arrives at a final set of constraints
or / and spescifies some of $ \nu_{\a}$. such constraints are
divided into two types: first-class constraints which have
vanishing Poisson brackets with all other constraints and
second-class constraints which have non-vanishing Poisson
brackets. As there is an even number of second-class constraints,
these can be used to eliminate conjugate pair of $(p's ,q's)$ and
$({\pi} 's ,{\bar {q}}'s)$ from the theory by expressing them as
functions of the remaining $(p's ,q's)$ and $({\pi} 's
,{\bar{q}}'s)$. The Dirac Hamiltonian for the remaining variables
is then the canonical Hamiltonian plus all the independent first
class constraints $\Psi_{\la}$. So that the total Hamiltonian is
defined as
\begin{equation}
H_T = H_c +\nu_{\la} \Psi_{\la}.
\end{equation}

Since first-class constraints are the generators of gauge
transformations, this will lead to the gauge freedom. In other
words, the equations of motion are still degenerate and depend on
the functional arbitrariness. Besides, some $\nu_{\la}$ are still
undermined. To remove this arbitrariness, one has to impose
external gauge fixing conditions for each first class constraints.

Fixing a gauge is not always an easy task, which make one be
careful when applying Dirac's method.

Now we would like to give the path integral formulation using the
canonical path integral method and demonstrate the fact that the
gauge fixing problem is solved naturally if this method is used.

\section{The canonical path integral formalism for second order Lagrangians}

Recently the canonical method [12-14] has been developed to
investigate singular systems using the Caratheodoy's equivalent
Lagrangian method and the equations of motion are obtained as
total differential equations in many variables.

Now we will give a brief review of the Caratheodory's equivalent
Lagrangian method. Let us consider a Lagrangian $L(q_i,
\dot{q_i}, \ddot{q_i}, t)$, we can obtain a completely equivalent
one by
\begin{equation}
L' = L(q_i, \dot{q_i}, \ddot{q_i},t ) - \frac{dS(q_i, \dot{q_i},
t)}{dt},
\end{equation}
such a function $S(q_i, \dot{q_i}, t)$ must satisfy

\bn &&\frac{\p S}{\p t}= - H_0,\\
&&H_0 = p_i \bar{q_i} +\pi_i \ddot{q_i} - L,\\
&&p_i = \frac{\p S}{\p q_i},\\
&&\pi_i = \frac{\p S}{\p \bar q_i}. \en These are the fundamental
equations of equivalence Lagrangian method.

If the rank of the Hess matrix $\frac{\p^2 L}{\p \ddot{q_i}\p
\ddot{q_j}}$ is $n-R$, $R < n$, then the generalized momenta
conjugated to the generalized coordinates $\bar{q_i}$ are defined
as

\bn &&\pi_a =\frac{\p L}{\p \dot{\bar{q_a}}},\;\;\;a= R+1,...,n,\\
&&\pi_{\a} =\frac{\p L}{\p \dot{\bar{q_{\a}}}}\;\;\;
\a=1,...,R.\en

Since the rank of the hess matrix $\frac{\p^2 L}{\p
\dot{\bar{q_i}}\p \dot{\bar{q_j}}}$ is $n-R$, then one can solve
the $n-R$ accelerations $\dot{\bar{q_{a}}}$ in terms of
coordinates $(q_{i}, \bar{q_{i}})$, the momenta $\pi_{a}$ and
$\dot{\bar{q_{\a}}}$ as follows
\begin{equation}
\dot{\bar{q_{a}}}= w_{a}(q_{i}, \bar{q_{i}}, \pi_{a},
\dot{\bar{q_{\a}}}).
\end{equation}
Substituting (18) in (17) one has
\begin{equation}
\pi_{\a} =\frac{\p L}{\p \dot{\bar{q_{\a}}}}|_{\dot{\bar{q_{a}}}=
w_{a}(q_{i}, \bar{q_{i}}, \pi_{a}, \dot{\bar{q_{\a}}})}=
-H_{\a}^{\pi}(q_{i}, \bar{q_{i}}, p_{b}, \pi_{a}).
\end{equation}

On the other hand, from equation (3) if the rank of the Hess
matrix $\frac{\p^2 L}{\p {\bar{q_i}}\p {\bar{q_j}}}$ is $n-r$, we
can obtain a similar expression for the momenta $p_{\g}$:
\begin{equation}
p_{\g} = -H_{\g}^{p}(q_{i}, \bar{q_{b}}, p_{b}, \pi_{a}),\;\;\;
\g=1,...,r,\;\; b = r+1,...,n.
\end{equation}

The Hamiltonian $H_{0}$ is defined as

\bn H_{0}=&& p_{b}\bar{q_{b}} + \bar{q_{\g}}p_{\g}|_{p_{\eps}=
-H_{\eps}^{p}} + \pi_{a} w_{a}\\&& + \bar{q_{\a}}
\pi_{\a}|_{p_{\bt}= -H_{\bt}^{p}} - L(q_{i}, \bar{q_{i}},
\dot{\bar{q_{\a}}}, \dot{\bar{q_{a}}}= w_{a}),\;\eps =1,...,r. \en

Relabeling the coordinates $t$ and $q_{\g}$ as $t_{0}$ and
$t_{\g}$ respectively, and $ \bar{q_{\a}}$ will be called
$t_{\a}$. defining the momenta $P_{0}$ as
\begin{equation}
P_{0} = \frac{\p S}{\p t},
\end{equation}
then the set of Hamilton Jacobi partial differential equations
[HJPDE] is expressed as

\bn &&H'_{0} = P_{0} + H_{0}(t_{0}, t_{\g}, t_{\a}; q_{b},
\bar{q_{a}}
; p_{b}= \frac{\p S}{\p q_{b}}; \pi_{a}= \frac{\p S}{\p
\bar{q_{a}}})=0,\\
&&{H'}_{\g}^{p} = p_{\g} + H_{\g}^{p}(t_{0}, t_{\g}, t_{\a};
q_{b}, \bar{q_{a}}; p_{b}= \frac{\p S}{\p q_{b}}; \pi_{a}=
\frac{\p S}{\p
\bar{q_{a}}})=0,\\
&&{H'}_{\a}^{\pi} = \pi_{\a} + H_{\a}^{\pi}(t_{0}, t_{\g},
t_{\a}; q_{b}, \bar{q_{a}}; p_{b}= \frac{\p S}{\p q_{b}}; \pi_{a}=
\frac{\p S}{\p \bar{q_{a}}})=0, \en

The equations of motion are obtained as total differential
equations in many variables as follows:

\bn &&dq_{i} = \frac{\p H'_{0}}{\p p_{i}}dt_{0} + \frac{\p
{H'}_{\g}^{p} }{\p p_{i}}dt_{\g}+ \frac{\p {H'}_{\a}^{\pi} }{\p
p_{i}}dt_{\a},\\
&&d\bar{q_{i}} = \frac{\p H'_{0}}{\p \pi_{i}}dt_{0} + \frac{\p
{H'}_{\g}^{p} }{\p \pi_{i}}dt_{\g}+ \frac{\p {H'}_{\a}^{\pi} }{\p
\pi_{i}}dt_{\a},\\
&&dp_{i} = -\frac{\p H'_{0}}{\p q_{i}}dt_{0} - \frac{\p
{H'}_{\g}^{p} }{\p q_{i}}dt_{\g} - \frac{\p {H'}_{\a}^{\pi} }{\p
q_{i}}dt_{\a},\\
&&d\pi_{i} = -\frac{\p H'_{0}}{\p \bar{q_{i}}}dt_{0} - \frac{\p
{H'}_{\g}^{p} }{\p \bar{q_{i}}}dt_{\g} - \frac{\p {H'}_{\a}^{\pi}
}{\p \bar{q_{i}}}dt_{\a},\\
&&dP_{0} = -\frac{\p H'_{0}}{\p t_{0}}dt_{0} - \frac{\p
{H'}_{\g}^{p} }{\p t_{0}}dt_{\g} - \frac{\p {H'}_{\a}^{\pi} }{\p
t_{0}}dt_{\a},\\
&&dZ = ( -H_{0} + p_{b}\frac{\p H'_{0}}{\p p_{b}} +
\pi_{a}\frac{\p H'_{0}}{\p \pi_{a}})dt_{0}\nonumber\\&&+ (
-H_{\g}^{p} + p_{b}\frac{\p {H'}_{\g}^{p} }{\p p_{b}} +
\pi_{a}\frac{\p {H'}_{\g}^{p} }{\p \pi_{a}})dt_{\g}\nonumber\\
&&+ ( -H_{\a}^{\pi} + p_{b}\frac{\p {H'}_{\a}^{\pi} }{\p p_{b}} +
\pi_{a}\frac{\p {H'}_{\a}^{\pi} }{\p \pi_{a}})dt_{\a}. \en The
set of equations (27-32) is integrable if [14]

\bn &&d{H'}_{0}=0,\\
&&d{H'}_{\g}^{p} =0,\\
&&d{H'}_{\a}^{\pi}= 0. \en If conditions (33-35) are not satisfied
identically, one considers them as new constraints and again
tests the consistency conditions. Thus, repeating this procedure
one may obtain a set of conditions. Hence, the canonical
formulation leads us to obtain the set of canonical phase - space
coordinates as follows

\bn &&q_{b} \equiv q_{b}(t_{0}, t_{\g}, t_{\a}),\;\;p_{b}\equiv
p_{b}(t_{0}, t_{\g},
t_{\a}),\;\;\nonumber\\&&b=r+1,...,n,\;\g=1,...,r,\;\a=1,...,R.\\
&&\bar{q_{a}} \equiv \bar{q_{a}}(t_{0}, t_{\g},
t_{\a}),\;\;\pi_{a}\equiv \pi_{a}(t_{0}, t_{\g},
t_{\a}),\;\;\nonumber\\&&a=R+1,...,n,\;\g=1,...,r,\;\a=1,...,R.
\en Besides the canonical action integral is obtained in terms of
the canonical coordinates. ${H'}_{0}, {H'}_{\g}^{p}$ and
${H'}_{\a}^{\pi}$ can be interpreted as the infinitesimal
generators of canonical transformations given by parameters
$t_{0}, t_{\g}$ and $t_{\a}$ respectively. In this case, the path
integral representation may be written as [9-13]

\bn \langle q_{b}, \bar{q_{a}}, t_{\g}, t_{\a}| {q'}_{b},
\bar{{q'}_{a}}, {t'}_{\g}, {t'}_{\a}\rangle&&=\int
\prod_{b=1}^{r}dq^{b}~dp^{b}\prod_{a=1}^{R}d{\bar{q^{a}}}~d{\bar{\pi^{a}}}\times\nonumber\\
&&\exp i \{\int_{t_{\g}, t_{\a}}^{{t'}_{\g},{t'}_{\a}}( -H_{0} +
p_{b}\frac{\p H'_{0}}{\p p_{b}} + \pi_{a}\frac{\p H'_{0}}{\p
\pi_{a}})dt_{0}\nonumber\\&&+ ( -H_{\g}^{p} + p_{b}\frac{\p
{H'}_{\g}^{p} }{\p p_{b}} +
\pi_{a}\frac{\p {H'}_{\g}^{p} }{\p \pi_{a}})dt_{\g}\nonumber\\
&&+ ( -H_{\a}^{\pi} + p_{b}\frac{\p {H'}_{\a}^{\pi} }{\p p_{b}} +
\pi_{a}\frac{\p {H'}_{\a}^{\pi} }{\p \pi_{a}})dt_{\a}\}. \en

\section{Example}

In this section we will consider a singular system with a
Lagrangian density depend on the dynamical field variables:
${\cal L}= {\cal L}({\e}, \p_{\m}{\e}, \p_{\m}\p_{\n}{\e}).$ One
can obtain the Euler Lagrange equations of motion as follows
\begin{equation}
\frac{\p {\cal L}}{\p {\e}}- {\p}_{\m}(\frac{\p {\cal
L}}{\p({\p}_{\m}{\e})}) + {\p}_{\m}{\p}_{\n}(\frac{\p {\cal
L}}{\p({\p}_{\m}{\p}_{\n}{\e})})=0,\;\;\;\;\m,\;\n=0, 1, 2, 3.
\end{equation}
The momenta conjugated to ${\dot{\e}}$ and ${\ddot{\e}}$ are:

\bn && p= \frac{\p {\cal L}}{\p {\dot{\e}}}- 2{\p}_{k}(\frac{\p
{\cal L}}{\p({\p}_{k}{\dot \e})}) - {\p}_{0}(\frac{\p {\cal
L}}{\p{\ddot\e}}),\;\;\;\;k=1, 2, 3,\\
&&\pi= \frac{\p {\cal L}}{\p{\ddot\e}}. \en

Now will we will consider the Podolsky electrodynamics as a
constrained system with second order Lagrangian . The Lagrangian
density for such a system is given as
\begin{equation}
{\cal L} = -\frac{1}{4}F^{{\m}{\n}}F_{{\m}{\n}} +
a^{2}{\p}_{\la}F^{{\a}{\la}}{\p}_{\rho}F_{\a}^{\rho},\;\;\;\;\m,\;\n,\;\a,\;\rho
=0, 1, 2, 3,
\end{equation}
where
\begin{equation}
F_{{\m}{\n}}= {\p_{\m}}A_{\n} -  {\p_{\n}}A_{\m},
\end{equation}
and the metric convention $g_{\m\n} =diag(+1, -1, -1, -1)$. With
the dynamical variables chosen as $ A^{\m}$ and $\bar{A^{\m}}=
{\dot A^{\m}},$ the conjugated momenta are obtained as

\bn&&p_{\m}= - F_{0{\m}} - 2 a^{2}(\p_{k} \p_{\la}
F^{0{\la}}{\delta}_{\m}^{k} - \p_{0} \p_{\la} F_{\m}^{\la}),\\
&&\pi_{\m}= 2 a^{2}( \p_{\la} F^{0{\la}}{\delta}_{\m}^{0} -
\p_{\la} F_{\m}^{\la}). \en

The primary constraints are

\bn &&H'_{1}=\pi_{0}=0,\\
&&H'_{2} = p_{0}- \p^{k}\pi_{k}=0. \en The expressible velocities
${\dot {\bar{A^{i}}}}$ are obtained as
\begin{equation}
{\dot {\bar{A^{i}}}} = \frac{1}{2a^{2}}\pi^{i} + \p_{k} F^{ik} +
\p^{i} \bar{A_{0}}.
\end{equation}
The canonical Hamiltonian is given by
\begin{equation}
H_{0}= \int d^{3} x(p_{\m}\bar{A^{\m}}+ \pi_{\m}{\dot
{\bar{A^{\m}}}} - {\cal L}).
\end{equation}
Making use of equation (48), we get

\bn&& H_{0}= \int d^{3} x[\bar{A_{0}}\p^{i} \pi_{i} +
p_{i}{\bar{A^{i}}}
+\frac{1}{4a^{2}}{\pi_{i}}{\pi^{i}}{\p_{k}F^{ik}} +
\pi_{i}\p_{k}F^{ik}\nonumber\\&&+ \pi_{i}\p^{i}\bar{A_{0}} +
\frac{1}{4}F_{\m\n}F^{\m\n}+ \frac{1}{2}(\bar{A_{i}}
-{\p_{i}{A_{0}}})(\bar{A^{i}} -{\p^{i}{A^{0}}})\nonumber\\&&-
a^{2} (\p_{k}{\bar {A^{k}}}-\p_{k}\p^{k}A_{0})(\p_{i}{\bar
{A^{i}}} -\p_{i}\p^{i}A_{0})]. \en

Equations (46),(47) and (50) lead us to obtain the set of Hamilton
Jacobi partial differential equations [HJPDE] as follows

\bn&& H'_{0}= P_{0} + H_{0}; \;\;\;\; P_{0} = \frac{\p S}{\p t},\\
&&H'_{1}=\pi_{0}=0,\;\;\;\;\;\;\pi_{0} = \frac{\p S}{\p \bar{A^{0}}},\\
 &&H'_{2} = p_{0}- \p^{k}\pi_{k}=0.\;\;\;\;\;p_{0} = \frac{\p S}{\p {A^{0}}}, \en

The equations of motion are obtained as total differential
equations in many variables as follows

\bn dA^{i}=&& \frac{\p {H'_{0}} }{\p p_{i}}dt + \frac{\p {H'_{1}}
}{\p
p_{i}}d \bar{A^{0}} + \frac{\p {H'_{2}} }{\p p_{i}}d {A^{0}},\\
=&& \bar{A^{i}}dt,\\
d\bar{A^{i}} =&&\frac{\p H'_{0} }{\p \pi_{i}}dt + \frac{\p
H'_{1}}{\p
\pi_{i}}d \bar{A^{0}} + \frac{\p H'_{2} }{\p \pi_{i}}d {A^{0}},\\
=&& ( \frac{1}{2a^{2}}\pi^{i} + \p_{k} F^{ik} + \p^{i}
\bar{A_{0}}),\\
dp^{i} =&&-\frac{\p H'_{0} }{\p A_{i}}dt - \frac{\p H'_{1}}{\p
A_{i}}d \bar{A^{0}} - \frac{\p H'_{2} }{\p A_{i}}d {A^{0}},\\
=&&(- \p^{i} \p^{k} \pi_{k} + \p_{i} \p^{k} \pi^{i} -\p_{k}
F^{ki})dt,\\
dp^{0} =&&-\frac{\p H'_{0} }{\p A_{0}}dt - \frac{\p H'_{1}}{\p
A_{0}}d \bar{A^{0}} - \frac{\p H'_{2} }{\p A_{0}}d {A^{0}},\\
=&&(-\p_{i}F^{0i}- 2a^{2} \p^{i}\p_{i}\p_{k}F_{0}^{k})dt,\\
d\pi^{i} =&&-\frac{\p H'_{0} }{\p \bar{A_{i}}}dt - \frac{\p
H'_{1}}{\p
\bar{A_{i}}}d \bar{A^{0}} - \frac{\p H'_{2} }{\p \bar{A_{i}}}d {A^{0}},\\
=&& (- p^{i} - F^{0i} - 2a^{2} \p^{i}\p_{k}F^{0k})dt,\\
d\pi^{0} =&&-\frac{\p H'_{0} }{\p \bar{A_{0}}}dt - \frac{\p
H'_{1}}{\p
\bar{A_{0}}}d \bar{A^{0}} - \frac{\p H'_{2} }{\p \bar{A_{0}}}d {A^{0}},\\
=&& (-\p_{k} p^{k})dt,\\
dP^{0} =&&-\frac{\p H'_{0} }{\p t}dt - \frac{\p H'_{1}}{\p t}d
\bar{A^{0}} - \frac{\p H'_{2} }{\p t}d {A^{0}}=0. \en

To check whether the set of equations (54-66) is integrable or
not, let us consider the total variations of (51-53). In fact

\bn&&dH'_{0}= H'_{3}dA_{0} = (\p^{k}p_{k})dA_{0}=0,\\
&&dH'_{0}= -H'_{3}dt = (-\p^{k}p_{k})dt=0,\\
&&dH'_{0}= H'_{3}dt = (\p^{k}p_{k})dt=0. \en The total variation
of $H'_{3}$ is identically zero. Hence, the equations of motion
are integrable and the canonical phase space coordinates $(A^{i},
p^{i}, \bar{A^{i}}, \pi^{i})$ are obtained in terms of parameters
$(t, A^{0}, \bar{A^{0}})$. Besides the canonical action integral
is obtained in terms of the canonical variables as

\bn &&dz= \int d^{3}x[\frac{1}{2a^{2}}{\pi_{i}}{\pi^{i}}-
\frac{1}{4}F_{\m\n}F^{\m\n}- \frac{1}{2}(\bar{A_{i}}
-{\p_{i}{A_{0}}})(\bar{A^{i}} -{\p^{i}{A^{0}}})\nonumber\\&&-
a^{2} (\p_{k}{\bar {A^{k}}}-\p_{k}\p^{k}A_{0})(\p_{i}{\bar
{A^{i}}} -\p_{i}\p^{i}A_{0})]dt. \en

Making use of equation (38) and equation (70), the path integral
for the Podolsky electrodynamics is given as

\bn\langle A^{i}, \bar{A^{i}}, t, A^{0}, \bar{A^{0}}|{{A}^{i}}',
{\bar{A^{i}}}', t', {{A}^{0}}', {\bar{{A}^{0}}}'\rangle&& = \int
\prod_{i=1}^{3}d A^{i}~dp^{i}~d\bar{A^{i}}
~d\bar{\pi^{i}}\times\nonumber\\ &&\exp i\{\int
d^{3}x[\frac{1}{2a^{2}}{\pi_{i}}{\pi^{i}}-
\frac{1}{4}F_{\m\n}F^{\m\n}\nonumber\\&&- \frac{1}{2}(\bar{A_{i}}
-{\p_{i}{A_{0}}})(\bar{A^{i}} -{\p^{i}{A^{0}}})\nonumber\\&&-
a^{2} (\p_{k}{\bar
{A^{k}}}-\p_{k}\p^{k}A_{0})\times\nonumber\\&&(\p_{i}{\bar
{A^{i}}} -\p_{i}\p^{i}A_{0})]dt\}. \en The path integral
representation (71) is an integration over the canonical
phase-space coordinates $(A^{i}, p^{i}, \bar{A^{i}}, \pi^{i})$.

\section{conclusion}

We have obtained the path integral for singular systems with
second order Lagrandians. For the Podolsky electrodynamics
example since the integrability conditions $dH'_{0}=0, dH'_{1}=0$
and $dH'_{2}=0$ are satisfied the canonical phase-space
coordinates $(A^{i}, p^{i}, \bar{A^{i}}, \pi^{i})$ are obtained
in terms of parameters $(t, A^{0}, \bar{A^{0}})$ and the path
integral (71) is obtained directly  as an integration over the
canonical phase-space coordinates without using any gauge fixing
conditions. The Faddeev's Popov [15,16] method treatment for this
model needs gauge fixing conditions to arrive at the result (71).
The generalization of the present work for Lagrangians of order
higher than two is given in reference [11].


\begin{thebibliography}{widest-label}
\bibitem{1}PODOLSKY B. and SCHWED P., Rev. Mod. Phys.,
$\mathbf{20}$ (1948) 40.
\bibitem{2}NESTERENKO v.v., Phys. Lett. B, $\mathbf{327}$ (1994) 50.
\bibitem{3}OSTROGRADSKI M., Mem. Ac. St. Petersbourg, $\mathbf{1}$
(1850) 385.
\bibitem{4}DIRAC P. A. M., "{\it Lectures on Quantum Mechanics}", Belfer
 Graduate School of Science, Yehiva University (A cademic Press,
 New York) 1964.
\bibitem{5}DIRAC P. A. M., Can. J. Math., $\mathbf{2}$ (1950)  129.
\bibitem{6}MUSLIH S. I., Nuovo Cimento B, $\mathbf{115}$ (2000) 1.
\bibitem{7}MUSLIH S. I., Nuovo Cimento B, $\mathbf{115}$ (2000) 7.
\bibitem{8}MUSLIH S. I., Hadronic J. $\mathbf{23}$ (2000) 203.
\bibitem{9}MUSLIH S. I., {\it " Path Integral Quantization of a Relativistic
Charged Particle in an External Electromagnetic Field"} to appear
in Hadronic J.
\bibitem{10}MUSLIH S. I., El-ZALAN H. and El-SABAA F., Int. J. of Theor. Phys.,
$\mathbf{23}$ (2000) 2505.
\bibitem{11}MUSLIH S. I., Math-Ph$/$0009015
\bibitem{12}GULER Y., Nuovo Cimento B, $\mathbf{107 }$(1992) 1389.
\bibitem{13}GULER Y., Nuovo Cimento B, $\mathbf{107}$ (1992) 1143.
\bibitem{14}MUSLIH S. I. and GULER Y., Nuovo Cimento B, $\mathbf{113}$
(1998) 277.
\bibitem{15}FADDEEV L. D. and POPOV V. M., Phys. Lett.
B, $\mathbf{24}$ (1967) 29.
\bibitem{16}FADDEEV L. D., Theor. Math. Phys., $\mathbf{1}$ (1970) 1.

\end{thebibliography}
\end{document}